\documentclass[preprint,aps,prd,nofootinbib,preprintnumbers,amsmath,amssymb]{revtex4}

\newcommand\fft[2]{\frac{#1}{#2}}
\newcommand\ft[2]{{\textstyle\frac{#1}{#2}}}
\newcommand{\beq}{\begin{equation}}
\newcommand{\eq}{\end{equation}}
\newcommand{\beqa}{\begin{eqnarray}}
\newcommand{\eqa}{\end{eqnarray}}

\newcommand{\ala}{\alpha_1}
\newcommand{\alb}{\alpha_2}
\newcommand{\alc}{\alpha_3}

\begin{document}
\preprint{MCTP-09-51}
\preprint{MIFP-09-44}
\preprint{DAMTP-2009-72}

\title{Higher Derivative Corrections to $R$-charged Black Holes:
Boundary Counterterms and the Mass-Charge Relation}

\author{Sera Cremonini}
\email{sera@physics.tamu.edu}
\affiliation{Centre for Theoretical Cosmology, DAMTP, Centre for Mathematical Sciences,\\
University of Cambridge, Wilberforce Road, Cambridge, CB3 0WA, UK}
\affiliation{George and Cynthia Mitchell Institute for Fundamental Physics and Astronomy,
Texas A\&M University, College Station, TX 77843--4242, USA}

\author{James T.~Liu}
\email{jimliu@umich.edu}
\affiliation{Michigan Center for Theoretical Physics,
Randall Laboratory of Physics,
The University of Michigan,
Ann Arbor, MI 48109--1040, USA}

\author{Phillip Szepietowski}
\email{pszepiet@umich.edu}
\affiliation{Michigan Center for Theoretical Physics,
Randall Laboratory of Physics,
The University of Michigan,
Ann Arbor, MI 48109--1040, USA}


\begin{abstract}
We carry out the holographic renormalization of Einstein-Maxwell theory with
curvature-squared corrections.  In particular, we demonstrate how to
construct the generalized Gibbons-Hawking surface term needed to ensure
a perturbatively well-defined variational principle.  This treatment ensures
the absence of ghost degrees of freedom at the linearized perturbative
order in the higher-derivative corrections.  We use the holographically
renormalized action to study the thermodynamics of $R$-charged black holes
with higher derivatives and to investigate their mass to charge ratio in
the extremal limit.
In five dimensions, there seems to be a connection between the sign
of the higher derivative couplings required to satisfy
the weak gravity conjecture and that violating the shear viscosity to entropy
bound. This is in turn related to possible constraints on the central charges of the dual CFT,
in particular to the sign of $c-a$.
\end{abstract}

\maketitle

\section{Introduction}

Higher derivative corrections to pure Einstein gravity have seen renewed
interest with the development of the AdS/CFT correspondence. In particular,
they have played an important role in many of the thermodynamic and
hydrodynamic studies that have emerged from applications of AdS/CFT to
strongly coupled gauge theories.  Since the Einstein-Hilbert action is only
the leading term in the string theory expansion, higher derivative corrections
are natural from an effective field theory point of view.  In the gravitational
sector, such corrections generally take the form $(\alpha')^n R^{n+1}$, where
$R$ denotes schematically the Riemann tensor and its contractions.  On the field
theory side of the correspondence they describe finite 't Hooft coupling
$\lambda$ and finite $N$ corrections.

In theories that are maximally supersymmetric (e.g.~IIB theory in ten
dimensions), the first corrections do not enter until $\alpha'^{\,3} R^4$ order.
However, generically the first non-trivial terms appear at curvature-squared
level.  This has motivated numerous recent holographic studies with $R^2$ terms
parameterized by
\beq
\label{curv2}
e^{-1}\delta \mathcal{L}=\alpha_1 R^2 + \alpha_2 R_{\mu\nu}^2
+ \alpha_3 R_{\mu\nu\rho\sigma}^2 \, .
\eq
In the absence of matter fields, the Einstein equation takes the form
$R_{\mu\nu}=-(d-1)g^2g_{\mu\nu}$, where $g=1/L$ is the inverse AdS radius.
As a result, the $\alpha_1$ and $\alpha_2$ terms in (\ref{curv2}) may be
shifted away by an on-shell field redefinition of the form
\begin{equation}
g_{\mu\nu}
\to g_{\mu\nu}+\lambda_1[R_{\mu\nu}+(d-1)g^2g_{\mu\nu}]+\lambda_2
g_{\mu\nu}[R+d(d-1)g^2],
\end{equation}
for appropriate choices of $\lambda_1$ and $\lambda_2$.  In particular,
such a field redefinition allows (\ref{curv2}) to be replaced by the
well-known Gauss-Bonnet combination
\begin{equation}
e^{-1}\mathcal{L}_{GB}=\alpha_3(R^2 -4 R_{\mu\nu}^2 + R_{\mu\nu\rho\sigma}^2),
\end{equation}
which is the unique curvature-squared combination that nevertheless yields
equations of motion that are no higher than second derivative in the metric.

Many of the positive features of the Gauss-Bonnet combination, including
exact Gauss-Bonnet black hole solutions, have been exploited in recent
investigations of AdS/CFT hydrodynamics \cite{Brigante:2007nu,Brigante:2008gz}.
However, it is important to realize
that the $\alpha_1$ and $\alpha_2$ terms in (\ref{curv2}) are not always
unphysical once matter fields are turned on.  For example, in an
Einstein-Maxwell theory, shifting away the $\alpha_1$ and $\alpha_2$ terms in
(\ref{curv2}) would at the same time generate new mixed terms of the form
$RF^2$ and $R_{\mu\nu}F^{\mu\lambda}F^\nu{}_\lambda$.  This is especially
relevant in studies of $R$-charged backgrounds in five-dimensional gauged
supergravity, where the natural curvature-square correction arises as
the Weyl-tensor squared, as opposed to the Gauss-Bonnet combination
\cite{Hanaki:2006pj,Cremonini:2008tw}.

\subsection{Perturbative approach to higher-derivative terms}

The purpose of this paper is to revisit the holographic renormalization of
$R$-squared AdS gravity and to demonstrate the systematic construction
of both generalized Gibbons-Hawking surface terms and local boundary
counterterms in theories with higher derivatives.  It is well known that
higher derivative theories generically lead to unpleasant features such as
ghosts and additional propagating degrees of freedom.  However, since the
theories we are interested in arise from the low energy limit of string
theory, it is only consistent to treat the higher derivative terms
perturbatively, as part of the $\alpha'$ expansion.  In this way, these
terms will not generate additional ghost modes, and thus will not drastically
alter the dynamics of the lowest order two-derivative theory.

As an example of what we mean by the perturbative treatment of higher
derivative terms, consider a toy model of a simple harmonic oscillator with
a four-derivative addition \cite{Simon:1990ic}
\begin{equation}
\label{eq:toy}
L=\ft12\dot x^2-\ft12\omega^2x^2-\ft12\alpha(\ddot x^2-\omega^2\dot x^2).
\end{equation}
The resulting equation of motion is
\begin{equation}
\label{eq:toyeom}
(1+\alpha\omega^2)\ddot x + \omega^2 x^2+\alpha x^{(4)}=0,
\end{equation}
and has solution
\begin{equation}
x(t) = A_1e^{i\omega t}+A_2e^{-i\omega t}
+A_3e^{it/\sqrt\alpha}+A_4e^{-it/\sqrt\alpha} \, .
\end{equation}
The first two terms are conventional, while the last two arise because of the
higher derivative nature of the model.  This demonstrates that additional
degrees of freedom are present in this theory, and in particular it is no
longer sufficient to specify only two boundary conditions when constructing
the Green's function.  This is also clear when considering the variation of
the action
\begin{equation}
\label{eq:toyds}
\delta S=-\int_{t_1}^{t_2}[\mbox{EOM}]dt+\Bigl[((1+\alpha\omega^2)\dot x
+\alpha\dddot x)\delta x - \alpha \ddot x\delta\dot x\Bigr]_{t_1}^{t_2}.
\end{equation}
In order to have a well-defined variational principle, we must hold both
$x$ and $\dot x$ fixed at the endpoints of the time interval.

In general, for finite non-zero $\alpha$, there is no possibility of avoiding
the complications of the higher-derivative theory.  However, it is instructive
to consider the limit $\alpha\to0$.  In this case, it is clear that the
second solution, with frequency $1/\sqrt\alpha$, is not perturbatively
connected to the $\alpha=0$ theory.  Assuming the toy Lagrangian
(\ref{eq:toy}) arises from an $\mathcal O(\alpha)$ expansion of a more
complete theory, it is then clear that the second solution would never have
appeared in the full theory, and thus must be discarded for perturbative
consistency.  A simple way of arriving at the perturbative solution is to
rewrite the equation of motion (\ref{eq:toyeom}) as
\begin{equation}
\ddot x+\omega^2 x^2=-\alpha\fft{d^2}{dt^2}(\ddot x+\omega^2 x),
\end{equation}
We may then substitute in the lowest order equation of motion to obtain
$\ddot x+\omega^2x^2=\mathcal O(\alpha^2)$, and in general iterate to any
arbitrary order of $\alpha$ (our choice of shifting the kinetic term in
(\ref{eq:toy}) leads to vanishing perturbative corrections in $\alpha$, but
in general they could be present).

While perturbative solutions to the equation of motion are routinely
investigated, it is often equally important to construct a well-defined
variational principle at the perturbative level.  Looking at the toy model,
the difficulty here arises from the $-\alpha\ddot x\delta\dot x$ surface
variation in (\ref{eq:toyds}).  In general, no surface term exists that can
remove the dependence on $\delta\dot x$ on the boundary  (after all, this is
a four derivative theory).  However, at the perturbative level, we may use
the lowest order equation of motion to rewrite $-\alpha\ddot x\delta\dot x
=\alpha\omega^2 x\delta\dot x+\mathcal O(\alpha^2)$.  This variation can
then be canceled at $\mathcal O(\alpha)$ by adding a surface term of the form
\begin{equation}
\label{eq:toysgh}
S_{\rm surface}=\Bigl[-\alpha\omega^2x\dot x\Bigr]_{t_1}^{t_2}.
\end{equation}
In principle, this can be continued order by order in $\alpha$.

Using this toy model, we have motivated the fact that there is a consistent
perturbative treatment of higher derivative gravitational theories arising
out of string theory.  In particular, the gravitational analog of
(\ref{eq:toysgh}) is a generalized Gibbons-Hawking surface term, and this
was constructed in a particular case in \cite{Buchel:2004di} when
examining the effect of the IIB $R^4$ term on the shear viscosity to entropy
density ratio $\eta/s$ in $\mathcal N=4$ super-Yang-Mills theory.  The
construction in \cite{Buchel:2004di} was based on scalar channel fluctuations,
and hence focused on an effective scalar field theory.  Our present aim is
to extend this construction to the full gravity theory, and hence to demonstrate
that (perturbative) holographic renormalization of higher derivative gravity
theories is indeed consistent.

Allowing for a gauge field, we focus on the holographic renormalization of
$d$-dimensional Einstein-Maxwell theory with generic curvature-squared
corrections given by
\begin{equation}
\label{eq:lag}
e^{-1}\mathcal L=R-\ft14F^2+(d-1)(d-2)g^2+\alpha_1R^2+\alpha_2R_{\mu\nu}^2
+\alpha_3R_{\mu\nu\rho\sigma}^2.
\end{equation}
The bulk action from this Lagrangian must be supplemented by a set of
surface terms, whose goal is to ensure that the variational principle is
well defined.  In fact, when defined on a space with boundary, the
two-derivative Einstein-Hilbert action itself requires the addition of
the Gibbons-Hawking surface term to cancel boundary variations which would
otherwise spoil the variational principle.  The presence of higher
derivative corrections leads to additional boundary terms which need to be
canceled, and therefore requires the inclusion of an appropriate
generalization of the Gibbons-Hawking term.

For particular combinations of curvature corrections, the so-called Lovelock
theories where the equations of motion involve no higher than second
derivatives of the metric -- which include the Gauss-Bonnet combination as a special case -- proper boundary terms have already been constructed
\cite{Teitelboim,Myers:1987yn}.  However, for more general corrections,
we must treat the corrections perturbatively, and only in this case does the construction of a
generalized Gibbons-Hawking term become feasible\footnote{A similar construction has also been done for $F(R)$ theories of gravity in \cite{Dyer:2008hb} and also for more general higher derivative theories in \cite{Nojiri:2000gv,Nojiri:2001ae,Nojiri:2001aj,Cvetic:2001bk}.}.
We demonstrate below how this is done, and furthermore construct the set of
local counterterms removing the leading divergences from the action.
This generalizes the case of Gauss-Bonnet gravity, for which all the counterterms
needed to regularize the action were constructed in
\cite{Brihaye:2008kh,Astefanesei:2008wz,Brihaye:2008xu,Liu:2008zf}.

\subsection{$R$-charged black holes and the mass-charge relation}

For an application of the counterterm corrected action, we will look at
$R$-charged black hole thermodynamics.  In fact, one of the driving forces
behind the studies of AdS/CFT at finite temperature has been the close
resemblance of the laws of black hole physics with those of standard
thermodynamics.  To extract thermodynamic quantities from black hole
backgrounds one typically evaluates the on-shell action $I$ and the boundary
stress tensor, given by
\beq
T^{ab} = \frac{2}{\sqrt{-h}} \frac{\delta I}{\delta h_{ab}}\, ,
\eq
where $h_{ab}$ denotes the boundary metric.  The on-shell value of the
gravitational action may then be identified with the thermodynamic potential
$\Omega$ according to $I =\beta \Omega$, where in the grand canonical
ensemble
\beq
\Omega = E -TS -Q_I \Phi^I \, .
\eq
Here $Q_I$ are a set of conserved $R$-charges and $\Phi^I$ their respective
potentials.  Holographic renormalization ensures that both $\Omega$ and
$E$ are finite in the above expression.

Below we will perturbatively construct the $d$-dimensional spherically
symmetric $R$-charged black hole solutions to the $R$-squared theory
(\ref{eq:lag}) and study their thermodynamic properties using the
holographically renormalized action.  Extracting the higher curvature effects
on the black hole mass will also allow us to discuss the weak gravity
conjecture in the context of AdS black holes.
In fact, according to the conjecture,
the linear mass-charge relation for extremal (not necessarily SUSY) black holes cannot
be exact, but should receive corrections as the charge decreases.
For extremal $R$-charged black-holes, we find a deviation from the leading
relation $m=q$ of the form
\beq
\frac{m}{q}=\left(\frac{m}{q}\right)_0
\biggl[1-\frac{1}{r_+^2}\biggl(\alpha_1 f_1(r_+) +\alpha_2 f_2(r_+)+\alpha_3 f_3(r_+)\biggr) \biggr]\, ,
\eq
where $r_+$ is the horizon radius, and the $f_i(r_+)$ are
all positive functions.
Thus, $m/q$ will necessarily decrease when all the couplings
$\alpha_i$ are positive. Clearly, it is still possible for the ratio to decrease if some
of the $\alpha_i$ are negative, and in this respect it is important to be able to determine
the precise form of the couplings from UV physics.

A feature which we would like to emphasize is that the deviation from the
$m=q$ relation seems to be tied  to the correction to some of the transport
coefficients which have been computed holographically in the  context of the
quark gluon plasma. In particular, the sign of the correction to the
shear viscosity to entropy ratio $\eta/s$ has received a lot of attention,
precisely because curvature-squared  terms have been shown to lead to a
violation of the KSS bound \cite{Kovtun:2003wp}.  For the examples that have
been studied thus far, the sign of the higher derivative couplings responsible
for the bound violation is precisely the same as that needed by the weak
gravity conjecture.  For instance, for the special case of Weyl-squared
corrections, where
$\alpha_1=\frac{1}{6}\, \alpha$, $\alpha_2= -\frac{4}{3}\, \alpha$,
$\alpha_3=\alpha$,  the mass-charge relation becomes
\beq
\frac{m}{q}=\left(\frac{m}{q}\right)_0
\biggl[1-\alpha \,\frac{f(r_+)}{r_+^2} \biggr]\, ,
\eq
where the function $f(r_+)$ is positive,
while the expression for $\eta/s$ takes the form
\beq
\frac{\eta}{s}=\frac{1}{4\pi}\biggl[1-\alpha \,g(Q) \biggr]\, ,
\eq
where $g(Q)$ is a non-negative function of the $R$-charge.

The outline of the paper is as follows.  Section II is dedicated to the
construction of the perturbative generalization of the Gibbons-Hawking surface
term for the $R^2$ action (\ref{eq:lag}).  Following this, in Section III
we present the local counterterms needed to render this action finite in
dimensions $d\le7$.  We then present the $R$-charged black hole solution in
Section IV and explore their thermodynamics in Section V, where we also
discuss the implications of the mass to charge ratio for the weak gravity
conjecture.

\section{Generalizing the Gibbons-Hawking surface term}

Before considering the higher derivative gravitational action, it is worth
recalling that the ordinary Einstein-Hilbert action
\begin{equation}
\label{0action}
S_{\rm bulk} = -\frac{1}{2\kappa_d^2}\int_{\mathcal M}
d^dx \sqrt{-g}R
\end{equation}
contains explicitly second derivatives of the metric $g_{\mu\nu}$.
Thus, on a space with a boundary, variation with respect to the metric
yields, in addition to the standard $\delta g_{\mu\nu}$ factors,
terms involving the normal derivative of the metric.  In order to have a
well-defined variational principle where the metric, but not its derivative,
is held fixed at the boundary, the Einstein-Hilbert action must be
supplemented by the Gibbons-Hawking surface term
\begin{equation}
S_{\rm GH}=- \frac{1}{\kappa_d^2}  \int_{\partial\mathcal M}
d^{d-1}x\sqrt{-h}K \, .
\label{eq:ghst}
\end{equation}
Here $K$ denotes the trace of the extrinsic curvature tensor, $K_{\mu \nu} = \nabla_{(\mu}n_{\nu)}$, where $n_{\mu}$ specifies the normal direction to
the boundary surface, and $h_{ab}$ is the boundary metric.  With the inclusion
of the Gibbons-Hawking term, the unwanted normal derivative terms are canceled,
and the variational principle is well-defined.

We now consider the addition of curvature-squared terms, and take the bulk
action to be of the form
\begin{equation}
S_{\rm bulk}=-\fft1{2\kappa_d^2}\int_{\mathcal M}
d^{d}x\sqrt{-g}\, \Bigl[R-\fft14F^2+(d-1)(d-2)g^2 + \alpha_1R^2
+\alpha_2R_{\mu\nu}^2+\alpha_3R_{\mu\nu\rho\sigma}^2\Bigr].
\label{eq:bulk1}
\end{equation}
In general, this four-derivative action gives rise to higher order equations
of motion.  However, for the special choice of coefficients $\alpha_1=\alpha_3$
and $\alpha_2=-4\alpha_3$, the higher derivative terms combine to form the
well-known Gauss-Bonnet term $R^2-4R_{\mu\nu}^2+R_{\mu\nu\rho\sigma}^2$, which
is the unique combination that gives rise to equations of motion involving no
higher than second derivatives of the metric.  This motivates us to rewrite
(\ref{eq:bulk1}) in the equivalent form
\begin{eqnarray}
&&S_{\rm bulk}=-\fft1{2\kappa_d^2}\int_{\mathcal M}
d^{d}x\sqrt{-g}\, \Bigl[R-\frac{1}{4}F^2+(d-1)(d-2)g^2
\nonumber \\
&& \kern14em+\tilde\alpha_1R^2+\tilde\alpha_2R_{\mu\nu}^2+\alpha_3
(R^2-4R_{\mu\nu}^2+R_{\mu\nu\rho\sigma}^2)\Bigr],
\label{eq:bulk2}
\end{eqnarray}
where
\begin{equation}
\tilde\alpha_1=\alpha_1-\alpha_3,\qquad\tilde\alpha_2=\alpha_2+4\alpha_3.
\label{tilde}
\end{equation}

For the special case of Gauss-Bonnet gravity, where $\tilde\alpha_1
=\tilde\alpha_2=0$, the Gibbons-Hawking surface term can be generalized
\cite{Teitelboim,Myers:1987yn}, and takes the form
\begin{eqnarray}
\label{GBGH}
S_{\rm GH}^{\mbox{\scriptsize Gauss-Bonnet}}&=&-\frac{1}{\kappa_d^2}
\int_{\partial\mathcal M} d^{d-1}x\sqrt{-h} \,
\alpha_3\Big[ -\frac{2}{3}K^3 + 2 KK_{ab}K^{ab}
-\frac{4}{3}K_{ab}K^{bc}K_{c}{}^{a} \nonumber \\
& & \hspace{2in} - 4(\mathcal{R}_{ab} -
\frac{1}{2}\mathcal{R}h_{ab})K^{ab} \Big],
\end{eqnarray}
where $\mathcal{R}_{ab}$ is the boundary Ricci tensor.  However, no equivalent
term exists for $\tilde\alpha_1$ and $\tilde\alpha_2$ non-vanishing, because
in this case the equations of motion are of higher order, and in general
it is no longer sufficient to specify only the metric (and not derivatives)
on the boundary.

This issue is unavoidable whenever we are faced with higher order equations
of motion.  However, we are really only interested in viewing the higher
order terms as corrections to the two-derivative action.  In this case, we
only need to develop a perturbative expansion where the higher derivative terms
do not generate their own dynamics, but instead contribute merely correction
terms, thus effectively maintaining a two-derivative equation of motion.  In
this case, it should be possible to write down an effective Gibbons-Hawking
term, not just for the Gauss-Bonnet combination, but also for the $R^2$ and
$R_{\mu\nu}^2$ terms in the action.  This has been done for $R^2$ corrections
in $d=5$ by introducing auxiliary fields \cite{Cvetic:2001bk}. However, one
can avoid the complications involved in utilizing auxiliary fields by working
directly with the perturbative expansion.

To see how this may be done, we begin with the observation that the ordinary Gibbons-Hawking term
(\ref{eq:ghst}) is designed to cancel the appropriate part of the
variation of the Einstein-Hilbert term, namely $\sqrt{-g}g^{\mu\nu}\delta
R_{\mu\nu}$.  With this in mind, consider the variation
\begin{eqnarray}
\delta[R+\tilde\alpha_1R^2+\tilde\alpha_2R_{\mu\nu}^2]
&=&\delta R+2\tilde\alpha_1R\delta R+2\tilde\alpha_2
(R^{\mu\nu}\delta R_{\mu\nu}+R_{\mu\rho}
R^\mu_{\;\;\sigma}\delta g^{\rho\sigma})\nonumber\\
&=&(g^{\mu\nu}+2\tilde\alpha_1Rg^{\mu\nu}+2\tilde\alpha_2R^{\mu\nu})
\delta R_{\mu\nu}\nonumber\\
&&
+(R_{\mu\nu}+2\tilde\alpha_1RR_{\mu\nu}+2\tilde\alpha_2R_{\mu\rho}
R_{\nu}^{\;\;\rho})\delta g^{\mu\nu}.
\end{eqnarray}
Substituting in the lowest order equation
\begin{equation}
R_{\mu\nu}=-(d-1)g^2g_{\mu\nu} + \frac{1}{2}\left(
F_{\mu\lambda}F_\nu{}^\lambda- \frac{1}{2(d-2)}g_{\mu\nu}F^2\right)
+\mathcal O(\alpha_i)
\end{equation}
results in
\begin{eqnarray}
\delta[R+\tilde\alpha_1R^2+\tilde\alpha_2R_{\mu\nu}^2]
&=&(1-2\tilde\alpha_1g^2d(d-1)-2\tilde\alpha_2g^2(d-1))g^{\mu\nu}\delta R_{\mu\nu}  \nonumber \\
&& + \fft1{2(d-2)}(\tilde{\alpha}_1(d-4) - \tilde{\alpha}_2)F^2
g^{\mu\nu}\delta R_{\mu\nu} + \tilde\alpha_2F^{\mu\lambda}F^{\nu}{}_\lambda\delta R_{\mu\nu}
+\cdots,\nonumber\\
\label{Rvar}
\end{eqnarray}
where we have ignored higher order terms as well as terms not related to the
variation $\delta R_{\mu\nu}$.

For the terms in (\ref{Rvar}) involving simply $g^{\mu\nu}\delta R_{\mu\nu}$,
it is straightforward to generalize the usual Gibbons-Hawking term,
(\ref{eq:ghst}), to obtain a corresponding surface term canceling the
variation of the normal derivative of the metric
\begin{eqnarray}
S^{1}_{\rm GH}&=&-\frac{1}{\kappa_d^2}\int_{\partial\mathcal M} d^{d-1}x
\sqrt{-h}\Bigl[(1- 2\tilde{\alpha}_1 g^2d(d-1) - 2\tilde{\alpha}_2 g^2(d-1))K \nonumber \\
&& \kern15em + \fft1{2(d-2)}(\tilde{\alpha}_1(d-4)-\tilde{\alpha}_2)KF^2\Bigr].
\label{eq:R2gh}
\end{eqnarray}
However, the last term in (\ref{Rvar}) is a not as straightforward to deal
with, and the variation $\delta R_{\mu\nu}$ must be computed explicitly.
We find,
\begin{eqnarray}
\int_{\mathcal M} d^dx \sqrt{-g} F^{\mu\lambda}F^\nu{}_\lambda
\delta R_{\mu\nu} \kern-6em&&\nonumber\\
&=&\int_{\mathcal M} d^dx \sqrt{-g}
F^{\mu\lambda}F^\nu{}_\lambda \left(\nabla_\sigma
\delta\Gamma^{\sigma}_{\mu\nu} - \nabla_\mu
\delta\Gamma^\sigma_{\nu\sigma}\right) \nonumber \\
&=&\frac{1}{2}\int_{\mathcal M} d^{d}x \sqrt{-g}
F^{\mu\lambda}F^\nu{}_\lambda\left(2n^\rho\nabla_{(\mu}\delta g_{\nu)\rho}
- n^\rho \nabla_\rho \delta g_{\mu\nu}
- n_\mu g^{\rho\sigma}\nabla_{\nu}\delta g_{\rho\sigma}\right) \nonumber \\
&=&\fft12\int_{\mathcal M}d^dx\sqrt{-g}\,[\mbox{bulk}]
+\frac{1}{2}\int_{\partial\mathcal M} d^{d-1}x \sqrt{-h}
\bigl(-h^a{}_ch^b{}_dF^{c\lambda}F^d{}_\lambda\nonumber\\
&&\kern15em -h^{ab}n_\mu F^{\mu\lambda} n_\nu F^\nu{}_\lambda\bigr)
n^\rho\nabla_\rho\delta g_{ab}+\cdots,\qquad
\end{eqnarray}
where in the last line we have kept only the terms on the boundary coming
from integration by parts and including normal derivatives of the metric.
The proper Gibbons-Hawking boundary term associated with this variation is
then simply:
\begin{equation}
S^{2}_{\rm GH}=-\frac{1}{\kappa_d^2}\int_{\partial\mathcal M} d^{d-1}x
\sqrt{-h}\frac{\tilde{\alpha}_2}{2}
\left(Kn_{\mu}F^{\mu\lambda}n_{\nu}F^\nu{}_\lambda
+ K_{ab}F^{a \lambda}F^{b}{}_{\lambda}\right).
\label{eq:F2gh}
\end{equation}
It is now clear that the full effective Gibbons-Hawking term generalizing
(\ref{eq:ghst}) is just the sum of (\ref{eq:R2gh}) and (\ref{eq:F2gh}),
which handles the $\tilde\alpha_1$ and $\tilde\alpha_2$ terms, and
(\ref{GBGH}), which takes care of the $\alpha_3$ Gauss-Bonnet combination:
\begin{eqnarray}
\label{fullGHwF}
S_{\rm GH}&=&-\frac{1}{\kappa_d^2}\int_{\partial\mathcal M}d^{d-1}x
\sqrt{-h}\Bigl[\bigl(1- 2\tilde{\alpha}_1 g^2d(d-1)
- 2\tilde{\alpha}_2 g^2(d-1)\bigr)K \nonumber \\
&&\kern4em + \fft1{2(d-2)}\bigl(\tilde{\alpha}_1(d-4) - \tilde{\alpha}_2\bigr)
KF^2 + \frac{\tilde{\alpha}_2}{2}\left(Kn_{\mu}F^{\mu\lambda}n_{\nu}
F^\nu{}_\lambda + K_{ab}F^{a \lambda}F^{b}{}_{\lambda}\right) \nonumber \\
&&\kern4em - 2 \alpha_3 \Bigl(\frac{1}{3}K^3 -
KK_{ab}K^{ab} +\frac{2}{3}K_{ab}K^{bc}K_{c}{}^{a} +
2(\mathcal{R}_{ab} - \frac{1}{2}\mathcal{R}h_{ab})K^{ab}
\Bigr)\Bigr].
\end{eqnarray}

We note that the Gibbons-Hawking term now involves the gauge field strength
evaluated on the boundary. Variation of $S_{\rm GH}$ then results in
$\delta F$ terms on the boundary, thus complicating the variational principle
for the potential $A_\mu$.  This can in principle be avoided by working in
the canonical ensemble, where the charge is held fixed, and which
corresponds to taking $\delta(n_\mu F^{\mu a})=0$ instead of $\delta A_\mu=0$
on the boundary.  A natural way to do this is to add a Hawking-Ross boundary
term of the form $\int_{\partial\mathcal M} d^{d-1}x \sqrt{-h} \, n_\mu
F^{\mu a}A_a$ to cancel the boundary term which arises from the variation
of the gauge kinetic term in the bulk action \cite{Hawking:1995ap}.
However, for our present purposes, all terms involving the field strength in (\ref{fullGHwF}) are actually subdominant and, in fact,  vanish for all of the
thermodynamic quantities discussed below. Therefore, we will chose to work in
the grand-canonical ensemble without adding the Hawking-Ross term.

\section{Boundary Counterterms}

It is well known that the gravitational action (\ref{eq:bulk1}) evaluated
on the background solution is divergent. The divergences can be removed,
however, using the method of holographic renormalization, which involves
introducing appropriate boundary counterterms $S_{\rm ct}$ so that the full
action
\begin{equation}
\Gamma = S_{\rm bulk}+S_{\rm GH}-S_{\rm ct},
\end{equation}
remains finite on-shell.  This method has become quite standard in the
framework of AdS/CFT, since the boundary counterterms have a natural
interpretation as conventional field theory counterterms in the dual CFT.

Along with counterterms to remove divergences, one is also free to add an
arbitrary number of finite counterterms.  While such terms shift the values
of the action and boundary stress tensor, they are natural from the CFT point
of view, since they correspond to the freedom to change renormalization
prescriptions.  Their inclusion has played a key role, for example, in
resolving the puzzle of the unusual mass/charge relation
$M\sim \frac{3}{2}\mu+Q-\frac{1}{3}g^2 Q^2$ observed in \cite{Buchel:2003re}
for single $R$-charged black holes in AdS$_5$, in apparent conflict with the
BPS bound $M\geq Q$, saturated in this case when $\mu=0$. With the addition of
an appropriate finite counterterm, the expected linear relation
$M\sim \frac{3}{2}\mu + 3Q$ may be restored \cite{Liu:2004it}.  The finite
counterterms are also necessary for maintaining diffeomorphism invariance in
the renormalized theory, and may be unambiguously generated using the
Hamilton-Jacobi approach to boundary counterterms.

In order to explore the appropriate counterterm structure needed to
regulate the action (\ref{eq:bulk1}), we first note that it admits a vacuum
AdS solution with
\begin{equation}
R_{\mu\nu}=-(d-1)g_{\rm eff}^2g_{\mu\nu},
\end{equation}
where
\begin{equation}
g_{\rm eff}^2=g^2\left(1+\tilde\alpha_1g^2\fft{d(d-1)(d-4)}{d-2}
+\tilde\alpha_2g^2\fft{(d-1)(d-4)}{d-2}+\alpha_3g^2(d-3)(d-4)\right)
\label{eq:geff}
\end{equation}
is the shifted inverse AdS radius.  Writing the vacuum AdS metric as
\begin{equation}
ds^2=-(k+g_{\rm eff}^2r^2)dt^2+\fft{dr^2}{k+g_{\rm eff}^2r^2}
+r^2d\Omega_{d-2,k}^2,
\label{eq:adsmet}
\end{equation}
it is easy to see that $\sqrt{-g}\sim r^{d-2}$, and hence that the leading
divergence of the on-shell goes as $r_0^{d-1}$ where $r_0$ is an appropriate
cutoff.

The counterterm action for the theory (\ref{eq:bulk1}) may be expanded in
powers of the inverse metric $h^{ab}\sim 1/r_0^2$:
\begin{equation}
S_{\rm ct}=\fft1{2\kappa_d^2}\int_{\partial\mathcal M}d^{d-1}x\sqrt{-h}
\, [A+B\mathcal R+C_1\mathcal R^2+C_2\mathcal R_{ab}^2+C_3\mathcal R_{abcd}^2 +\cdots]\, .
\label{eq:sct}
\end{equation}
Note that we have ignored possible counterterms built out of $F_{\mu\nu}$
since in the configurations we are interested in the gauge field vanishes
sufficiently rapidly at the boundary so that it will not contribute to any
potential counterterms.  The $A$ and $B$ coefficients are chosen to cancel
the $r_0^{d-1}$ and $r_0^{d-3}$ power law divergences, respectively, while
the $C_i$ terms will cancel the $r_0^{d-5}$ divergence.  Note, however, that
at lowest order the asymptotic Einstein condition
$R_{\mu\nu}=-(d-1)g^2g_{\mu\nu}$ along with the boundary symmetry implied by
(\ref{eq:adsmet}) ensures that the boundary curvature satisfies the algebraic
relation $\mathcal R^2=(d-2)\mathcal R_{ab}^2$.  Furthermore, isotropy of the
transverse space relates $\mathcal R_{abcd}^2$ to the other boundary curvature
squared quantities as well.  What this means is that divergence cancellation
by itself is insufficient to fix the relative factors among the $C_i$
coefficients.

An elegant way around this ambiguity in fixing the $C_i$ coefficients
is to use the Hamilton-Jacobi method to obtain the counterterms.  In
particular, this was done in \cite{Liu:2008zf} to generate the counterterms
for the Gauss-Bonnet component of the action proportional to $\alpha_3$.
(These counterterms were previously constructed in
\cite{Brihaye:2008kh,Astefanesei:2008wz,Brihaye:2008xu} using more direct
methods.) In order to determine the $\tilde\alpha_1$ and $\tilde\alpha_2$
dependent counterterms, we may take a shortcut and note that they may be
absorbed by a field redefinition in the asymptotic limit.  In this case,
their only effect is to rescale the usual counterterms for the
two-derivative theory, which is proportional to the combination
\begin{equation}
\mathcal R_{ab}^2-\fft{d-1}{4(d-2)}\mathcal R^2
\end{equation}
at curvature squared order.  At the linear level, we combine the various
ingredients to obtain
\begin{eqnarray}
\label{eq:fullct}
S_{\rm ct}&=&\fft1{2\kappa_d^2}\int_{\partial\mathcal M}d^{d-1}x\sqrt{-h}
\biggl[2g(d-2)\biggl(1-\fft12\tilde\alpha_1g^2\fft{d(d-1)(3d-4)}{d-2}
\nonumber\\
&&\kern10em-\fft12\tilde\alpha_2g^2\fft{(d-1)(3d-4)}{d-2} -\fft16\alpha_3g^2(d-3)(d-4)\biggr)\nonumber\\
&&\kern4em+\fft1{g(d-3)}\biggl(1-\fft12\tilde\alpha_1g^2\fft{d(d-1)(5d-12)}{d-2}
\nonumber\\
&&\kern10em-\fft12\tilde\alpha_2g^2\fft{(d-1)(5d-12)}{d-2} +\fft32\alpha_3g^2(d-3)(d-4)\biggr)\mathcal R\nonumber\\
&&\kern4em+\fft1{g^3(d-3)^2(d-5)}\biggl(1-\fft12\tilde\alpha_1g^2
\fft{d(d-1)(7d-20)}{d-2}-\fft12\tilde\alpha_2g^2\fft{(d-1)(7d-20)}{d-2}
\nonumber\\
&&\kern14em-\fft72\alpha_3g^2(d-3)(d-4)\biggr)
\left(\mathcal R_{ab}^2-\fft{d-1}{4(d-2)}\mathcal R^2\right)\nonumber\\
&&\kern4em+\fft{\alpha_3}{g(d-5)}\left(\mathcal R^2-4\mathcal R_{ab}^2
+\mathcal R_{abcd}^2\right)+\cdots\biggr].
\end{eqnarray}
We have only explicitly worked out the counterterms up to
$\mathcal O(r_0^{d-5})$.  This is sufficient to cancel divergences for
$d\le7$, but is insufficient for removing finite terms that spoil
diffeomorphism invariance in $d=7$.  Hence our results are explicit
only for $d<7$, although the counterterm action can be extended to
arbitrary dimension if desired.

\section{The $R^2$ corrected black hole solution}

The full theory we are interested in is determined by the bulk action
(\ref{eq:bulk1}) along with the generalized Gibbons-Hawking term
(\ref{fullGHwF}) and counterterm action (\ref{eq:fullct}).  We now turn
to the construction of $R^2$ corrected spherically symmetric black hole
solutions to this system.
Since we are working to linear order in $\alpha_i$, we may substitute the
lowest order equations of motion wherever possible into the higher
curvature terms.  We find that the Einstein equation takes the form
\begin{eqnarray}
&&R_{\mu\nu} + (d-1)g^2g_{\mu\nu} + \frac{1}{4(d-2)}F^2g_{\mu\nu}  -
\frac{1}{2}F_{\mu\lambda}F_\nu{}^\lambda  = \nonumber \\
&& \kern3em \Biggl[ -g^4(\alpha_1d+\alpha_2)\frac{(d-4)(d-1)^2}{d-2}
-g^2\bigl(\alpha_1(d^2-8)+\alpha_2(3d-8)\bigr)
\frac{(d-1)}{2(d-2)^2}F^2 \nonumber \\
&& \kern4em +\bigl(\alpha_1(d-4)(3d-8)-\alpha_2(5d-12)\bigr)
\frac{1}{16(d-2)^3}(F^2)^2 + \frac{\alpha_2}{4(d-2)}
F_{\gamma\lambda}F^{\lambda\sigma}F_{\sigma\rho}F^{\rho\gamma} \nonumber \\
&& \kern4em +\bigl(\alpha_1(d-4)+\alpha_2(d-3)+\alpha_3(3d-8)\bigr)
\frac{1}{2(d-2)^2}\nabla_\lambda\nabla^\lambda F^2
+ \frac{\alpha_3}{d-2}R_{\gamma\rho\lambda\sigma}^2
\Biggr] g_{\mu\nu} \nonumber \\
&& \kern4em
+ g^2(\ala d+\alb-2\alc)(d-1)F_{\mu\lambda}F_\nu{}^\lambda+
\alpha_3F_{\mu\lambda}F^{\lambda\sigma}F_{\sigma\rho}F^\rho{}_\nu
\nonumber \\
&& \kern3em
-\bigl(\alpha_1(d-4)-\alpha_2+2\alpha_3)\frac{1}{4(d-2)}
F^2F_{\mu\lambda}F_\nu{}^\lambda-2\alpha_3
R_{\mu\rho\lambda\sigma}R_{\nu}{}^{\rho\lambda\sigma} \nonumber \\
&& \kern3em - (\alpha_2+2\alpha_3)R_{\mu\rho\nu\lambda}
F^{\rho\sigma}F^{\lambda}{}_\sigma - \frac{1}{2}(\alpha_2+4\alpha_3)
\nabla_\lambda\nabla^\lambda(F_{\mu\lambda}F_\nu{}^\lambda) \nonumber \\
&& \kern3em + (2\alpha_1+\alpha_2+2\alpha_3)\frac{(d-4)}{4(d-2)}
\nabla_\mu\nabla_\nu F^2\, ,
\end{eqnarray}
while the Maxwell equation is simply
\begin{equation}
\nabla^{\mu}F_{\mu\nu} = 0.
\end{equation}
The presence of $F^4$ terms in the Einstein equation indicates that we will
end up with metric terms up to $\mathcal O(Q^4)$ where $Q$ is the electric
charge.

We now take the spherically symmetric metric ansatz
\begin{equation}
ds^2=-f_1(r) \, dt^2+\frac{1}{f_2(r)} \, dr^2+ \, r^2d\Omega_{d-2,k}^2 \, ,
\label{eq:bhmet}
\end{equation}
where $k=1,0,-1$ specifies the curvature of the transverse space.
Inserting this into the Einstein equations yields the solution to
linear order in the $\alpha_i$:
\begin{eqnarray}
\label{eq:bhsoln}
f_1(r)\!&=&\! k + g_{\rm eff}^2r^2- \frac{\mu}{r^{d-3}}\nonumber \\
&&\! +\bigl(1+2g^2(\tilde\alpha_1d(d-1)-\tilde\alpha_2(d^2-6d+7)+\alpha_3(d-3)(d-4))\bigr) \frac{Q^2}{2(d-2)(d-3)r^{2(d-3)}}\! \nonumber  \\
&&\! + \frac{kQ^2}{r^{2d-4}}\left(2\tilde{\alpha}_1\frac{(d-4)}{(d-2)^2} - \tilde{\alpha}_2\frac{d^2-6d+10}{(d-2)^2}\right) + \alc(d-3)(d-4) \frac{\mu^2}{r^{2d-4}}\nonumber \\
&&\!- \frac{\mu Q^2}{r^{3d-7}}\left(\tilde{\alpha}_1\frac{(d-1)(d-4)}{(d-2)^2} - \frac{\tilde{\alpha}_2}{(d-2)^2} + \alc \frac{(d-4)}{(d-2)}\right)\nonumber \\
&&\! + \frac{Q^4}{4r^{4d-10}}\bigg(\tilde{\alpha}_1\frac{(d-4)(11d^2-45d+44)}{(d-2)^3(d-3)(3d-7)} + \tilde{\alpha}_2\frac{4d^3-33d^2+83d-64}{(d-2)^3(d-3)(3d-7)} \nonumber \\
&&\kern7em  +\alc\frac{(d-4)}{(d-2)^2(d-3)}\bigg) \nonumber \\
f_2(r) &=& \left( 1 - 2\gamma\frac{Q^2}{r^{2d-4}}\right)f_1(r) \, ,
\end{eqnarray}
where $g_{\rm eff}$ is defined in (\ref{eq:geff}) and
\begin{equation}
\gamma = \tilde{\alpha}_1\frac{(2d-3)(d-4)}{(d-2)^2}
+ \tilde{\alpha}_2\frac{d^2-5d+5}{(d-2)^2}.
\end{equation}
The gauge field is given by
\begin{equation}
A_t = \frac{Q}{(d-3)r^{d-3}} +\gamma\frac{Q^3}{(3d-7)r^{3d-7}},
\end{equation}
up to a possible constant.

Other than $k$, the black hole depends on two parameters: $\mu$, which is
related to the mass, and $Q$, which is essentially the electric charge.
Note that the mass parameter
$\mu$ is shifted from the conventional Gauss-Bonnet black hole mass parameter
by a constant proportional to $\alpha_3$.  In particular, the
Gauss-Bonnet theory ($\tilde\alpha_1=\tilde\alpha_2=0$) admits an exact
solution with a corresponding mass parameter $\hat\mu$ of the form
\begin{equation}
ds^2 = -f dt^2 + \frac{1}{f} dr^2 + r^2 d\Omega_{d-2,k}^2,
\end{equation}
where \cite{Boulware:1985wk,Wheeler:1985nh,Wiltshire:1985us,Cai:2001dz}
\begin{equation}
f=k+\fft{r^2}{2\tilde\alpha_3}\left[1\mp\sqrt{1+4\tilde\alpha_3
\left(\fft{\hat\mu}{r^{d-1}}-g^2 - \frac{Q^2}{2(d-2)(d-3)r^{2(d-2)}}\right)}\right],
\label{eq:egbbh}
\end{equation}
and $\tilde\alpha_3=\alpha_3(d-3)(d-4)$. Taking the `negative' branch of
(\ref{eq:egbbh}), which is the only one that admits a perturbative expansion,
we find to linear order in $\alpha_3$
\begin{eqnarray}
f&=&k+g_{\rm eff}^2r^2 - (1+2g^2\tilde\alc)\fft{\hat\mu}{r^{d-3}}
+(1+2g^2\tilde{\alpha}_3) \frac{Q^2}{2(d-2)(d-3)r^{2(d-3)}}\nonumber\\
&& +\tilde{\alpha}_3\fft{\hat\mu^2}{r^{2(d-2)}}
- \frac{\tilde{\alpha}_3\hat{\mu}Q^2}{(d-2)(d-3)r^{3d-7}}
+ \frac{\tilde{\alpha}_3 Q^4}{4(d-2)^2(d-3)^2r^{2(2d-5)}},
\end{eqnarray}
where in this case $g_{\rm eff}^2=g^2(1+g^2\tilde\alpha_3)$.
Comparing this with (\ref{eq:bhsoln}) demonstrates the relation
\begin{equation}
\mu=\hat\mu(1+2g^2\tilde\alpha_3)=\hat\mu\bigl(1+2g^2\alpha_3(d-3)(d-4)\bigr).
\end{equation}
Note also that for $Q=0$ the dependence of the solution (\ref{eq:bhsoln}) on
$\tilde\alpha_1$ and $\tilde\alpha_2$ is indirect through the shift in
$g_{\rm eff}$.  This is related to the fact that these contributions
may be removed at linear order through a field redefinition. However, with nonzero charge, a field redefinition of the form $g_{\mu\nu} \rightarrow g_{\mu\nu} + aRg_{\mu\nu} + b R_{\mu\nu}$ can in principle remove the $R^2$ and $R_{\mu\nu}^2$ terms in the action but also generates $RF^2$ and $R_{\mu\nu}F^{\mu\lambda}F^\nu{}_\lambda$ terms, implying that the coefficients $\tilde{\alpha}_1$ and $\tilde{\alpha}_2$ remain physical \cite{Liu:2008kt}.

\section{Thermodynamics}
\label{ThermoSection}

Given the holographically renormalized action, it is straightforward to study
the thermodynamics of the $R$-charged black holes.  We begin with the
temperature, which is given by the surface gravity of the black
hole, or equivalently by the requirement of the absence of a conical
singularity at the horizon of the Euclideanized black hole.  The relevant
part of the Euclideanized metric has the form
\begin{equation}
ds^2 = f_1(r)d\tau^2 + \fft{dr^2}{f_2(r)},
\end{equation}
where both $f_1$ and $f_2$ have a zero at the outer horizon, $f_1(r_+)=
f_2(r_+)=0$.  In this case, the temperature is given by
\begin{equation}
T=\fft1{4\pi}\left[\sqrt{f_1'(r)f_2'(r)}\right]_{r=r_+}.
\end{equation}
For $f_1$ and $f_2$ given in (\ref{eq:bhsoln}), we find:
\begin{eqnarray}
T &=& \fft1{4\pi r_+} \Biggl[(d-3)\fft\mu{r_{+}^{d-3}}+2g_{\rm eff}^2r_{+}^2- \frac{Q^2}{(d-2)r_+^{2d-6}}
-2\alpha_3(d-2)(d-3)(d-4)\fft{\mu^2}{r_+^{2d-4}}
  \nonumber \\
&&- \frac{g^2 Q^2}{r_+^{2d-6}} \bigg(2\tilde{\alpha}_1\frac{(d^4-36d^2+107d-84)}{(d-2)^2(d-3)}-2\tilde{\alpha}_2\frac{(d^4-14d^3+65d^2-121d+77)}{(d-2)^2(d-3)} \nonumber \\
&& \kern5em + 2\alc\frac{(d-3)(d-4)}{(d-2)}  \bigg) + \frac{k \, Q^2}{r_+^{2d-4}}\left(-12\tilde{\alpha}_1\frac{(d-4)}{(d-3)} +\tilde{\alpha}_2\frac{2(d-4)(d-5)}{(d-3)}\right) \nonumber  \\ &&+ \frac{\mu \, Q^2}{r_+^{3d-7}} \left(\tilde{\alpha}_1\frac{(d-4)(d^2+6d-15)}{(d-2)(d-3)}-\tilde{\alpha}_2\frac{(d^3-13d^2+49d-53)}{(d-2)(d-3)}
+\alc\frac{(d-4)(3d-7)}{(d-2)}\right)\nonumber \\
&&-  \frac{Q^4}{r_+^{4d-10}} \bigg( \tilde{\alpha}_1 \frac{(d-4)(10 d^3-49d^2+84d-57)}{2(d-2)^2(d-3)^2(3d-7)} + \tilde{\alpha}_2\frac{(2d^4-14d^3+31d^2-32d+25)}{2(d-2)^2(d-3)^2(3d-7)} \nonumber \\
&& \kern5em +\alc\frac{(2d-5)(d-4)}{2(d-2)^2(d-3)} \bigg) \Biggr] .
\label{eq:T0}
\end{eqnarray}
While this expression is written in terms of the parameters $\mu$, $r_+$
and $Q$, they are not all independent.  In particular, $\mu$ may be
written in terms of $r_+$ and $Q$ through the horizon condition $f_1(r_+)=0$
(although $\mu$ enters quadratically in (\ref{eq:bhsoln}), it is only necessary
to obtain $\mu$ to first order in the $\alpha_i$).

The entropy can be obtained by using Wald's formula
\begin{equation}
S=-2\pi\int_{\rm horizon}E^{\mu\nu\rho\sigma}\epsilon_{\mu\nu}
\epsilon_{\rho\sigma}d^{d-2}x,
\end{equation}
where
\begin{equation}
E^{\mu\nu\rho\sigma}=\fft{\delta S_{\rm bulk}}{\delta R_{\mu\nu\rho\sigma}}
\bigg|_{g_{\mu\nu}~\rm fixed},
\end{equation}
and $\epsilon_{\mu\nu}$ is the binormal to the horizon.
For the action (\ref{eq:bulk1}), we have
\begin{eqnarray}
E^{\mu\nu\rho\sigma}&=&-\fft1{2\kappa_d^2}\sqrt{-g}\Bigl[\ft12(1+\alpha_1R)
(g^{\mu\rho}g^{\nu\sigma}-g^{\mu\sigma}g^{\nu\rho})\nonumber\\
&&\kern5em+\ft12\alpha_2
(g^{\mu\rho}R^{\nu\sigma}+g^{\nu\sigma}R^{\mu\rho}
-g^{\mu\sigma}R^{\nu\rho}-g^{\nu\rho}R^{\mu\sigma})
+2\alpha_3R^{\mu\nu\rho\sigma}\Bigr],
\end{eqnarray}
in which case we find the entropy to be
\begin{eqnarray}
S&=&\fft{2\pi\omega_{d-2,k}}{\kappa_d^2}r_+^{d-2}\bigg[
1-2\tilde\alpha_1g^2d(d-1)-2\tilde\alpha_2g^2(d-1)+2\alpha_3(d-2)(d-3)
\fft{k}{r_+^2} \nonumber \\
&& \hspace{1.3in} -\frac{Q^2}{r_+^{2d-4}}\left(\tilde{\alpha}_1 \frac{d-4}{d-2}+ \tilde{\alpha}_2\frac{d-3}{d-2}\right)\bigg].
\label{eq:S0}
\end{eqnarray}
Here $\omega_{d-2,k}$ denotes the area of the transverse space given by
$d\Omega_{d-2,k}$.

The next ingredient we are interested in is the energy, which can be
extracted from the time-time component of the boundary stress tensor,
\begin{eqnarray}
T_{ab}&=& \frac{2}{\sqrt{-h}}\frac{\delta S}{\delta h^{ab}}\nonumber \\
&=&\frac{1}{2\kappa_d^2}\bigg[2\big(1- 2\tilde{\alpha}_1 g^2d(d-1) - 2\tilde{\alpha}_2 g^2(d-1)\big)(K_{ab} - Kh_{ab}) \nonumber \\
&&\kern2em + \left(\tilde{\alpha}_1\frac{(d-4)}{d-2}-\frac{\tilde{\alpha}_2}{d-2}\right)\left(F^2K_{ab} +2KF_{\lambda a}F^\lambda{}_b - \frac{1}{2}KF^2\right)   \nonumber \\
&&\kern2em + \tilde{\alpha}_2\left( K_{ab}h_{cd}n_\mu F^{\mu c}n_\nu F^{\nu d} + Kn_\mu F^{\mu}{}_a n_\nu F^\nu{}_b - \frac{1}{2}Kh_{cd}n_\mu F^{\mu c}n_\nu F^{\nu d}h_{ab}\right) \nonumber \\
&& \kern2em  + \tilde{\alpha}_2 \left(K_{cd}F^c{}_aF^d{}_b - \frac{1}{2}K_{cd}F^{c\lambda}F^d{}_\lambda h_{ab}\right)\bigg] + T^{GB}_{ab} + T^{CT}_{ab}\, , \nonumber \\
\end{eqnarray}
giving us the refreshingly simple expression
\begin{equation}
\label{eq:E}
E= \frac{\omega_{d-2}}{2\kappa_d^2}(d-2)\mu\left(1-2\tilde{\alpha}_1 g^2d(d-1) - 2\tilde{\alpha}_2 g^2(d-1) - 2 \alc g^2(d-3)(d-4)\right)\, ,
\end{equation}
which we expect to be valid in arbitrary dimension $d$.
Notice that in the absence of higher derivative corrections this expression
reproduces the familiar result $E\sim \mu$ found in \cite{Chamblin:1999tk}.
This also matches the Gauss-Bonnet black hole mass
\cite{Myers:1988ze,Cai:2001dz} in the case $\tilde\alpha_1=\tilde\alpha_2=0$,
and agrees with \cite{Cvetic:2001bk}, with arbitrary $\alpha_i$ coefficients
(note that we have removed the $k^2$ dependent `Casimir energy' by the addition
of finite counterterms, which was not done in \cite{Cvetic:2001bk}).

The final quantity we are interested in finding is the thermodynamic potential,
which can be obtained by evaluating the complete on-shell action:
\begin{equation}
\beta \Omega=S_{\rm bulk}+S_{\rm GH}+S_{\rm ct}.
\end{equation}
Computing this explicitly we find the renormalized free energy:
\begin{eqnarray}
\Omega&=&\fft{\omega_{d-2,k}}{2\kappa_d^2}\biggl[\mu\left(1-2\tilde\alpha_1g^2
d(d-1)-2\tilde\alpha_2g^2(d-1)-2\alpha_3g^2(d-2)(d-3)\right)\nonumber\\
&&-2g^2r_+^{d-1}\left(1-\tilde\alpha_1g^2\fft{d^2(d-1)}{d-2}
-\tilde\alpha_2g^2\fft{d(d-1)}{d-2}-\alpha_3g^2d(d-3)\right) \nonumber\\
&& - \frac{Q^2}{(d-2)(d-3)r_+^{d-3}} +\frac{kQ^2}{r_+^{d-1}}\left(4\tilde{\alpha}_1 \frac{(d-1)(d-4)}{(d-2)} + \tilde{\alpha}_2 \frac{d(d-4)}{(d-2)}\right) \nonumber \\
&& +\frac{g^2Q^2}{r_+^{d-3}}\left(2\tilde{\alpha}_1\frac{(d-1)(d-4)(2d-3)}{(d-2)^2} + \tilde{\alpha}_2\frac{d^3-4d^2+6}{(d-2)^2} -2\alc\frac{(d-4)}{(d-2)} \right) \nonumber \\
&& + \frac{\mu Q^2}{r_+^{2d-4}}\left(-4\tilde{\alpha}_1\frac{(d-1)(d-4)}{(d-2)} -\tilde{\alpha}_2\frac{d(d-4)}{(d-2)}+2\alc\frac{(2d-5)}{(d-2)}\right) -2\alpha_3(d-2)(d-3)\fft{\mu^2}{r_+^{d-1}} \nonumber \\
&& + \frac{Q^4}{2r_+^{3d-7}}\bigg(\tilde{\alpha}_1\frac{(d-4)(12d^2-45d+43)}{(d-2)^2(d-3)(3d-7)} + \tilde{\alpha}_2\frac{(3d^3-23d^2+53d-39)}{(d-2)^2(d-3)(3d-7)} \nonumber \\
&& \kern4em  - \alc\frac{(3d-8)}{(d-2)^2(d-3)}\bigg)    \biggr],
\label{eq:F0}
\end{eqnarray}
where we again recall that $\mu$ is a redundant parameter, and may be rewritten
in terms of $r_+$ and $Q$.

Since the $d$-dimensional expressions are rather unwieldy, we have checked our
calculations by verifying that the thermodynamic potential and energy satisfy
\begin{equation}
\Omega = E - TS - \mathcal{Q}\Phi,
\end{equation}
and the first law,
\begin{equation}
\label{firstlaw}
dE = TdS + \Phi d\mathcal{Q} \, .
\end{equation}
Here $\Phi$ is the chemical potential, defined as the difference in the potential between the horizon and spatial infinity,
\begin{equation}
\Phi(r_+) = A_t(r\rightarrow\infty) - A_t(r=r_+)\, ,
\end{equation}
and $\mathcal{Q} = (\omega_{d-2}/2\kappa_d^2) Q$ is the normalized electric charge which is unmodified by the higher derivative terms.

Finally, we note that a subtlety arises when applying the above thermodynamic
expressions in an AdS/CFT context.  For the $R$-charged black hole solution,
we have chosen a parameterization of the background which is asymptotic to
vacuum AdS given by (\ref{eq:adsmet}).  Taking $r\to\infty$, this has the
form
\begin{equation}
ds^2\sim -g_{\rm eff}^2r^2dt^2+r^2d\Omega_{d-2,k}^2
+\fft{dr^2}{g_{\rm eff}^2r^2}.
\end{equation}
Working on the Poincar\'e patch ($k=0$), the natural spatial coordinates
are written in terms of the zeroth order AdS radius, so that
\begin{eqnarray}
ds^2&\sim& -g_{\rm eff}^2r^2dt^2+g^2r^2d\vec x\,^2+\fft{dr^2}{g_{\rm eff}^2r^2}
\nonumber\\
&\sim&g^2r^2\left(-\fft{g_{\rm eff}^2}{g^2}dt^2+d\vec x\,^2\right)
+\fft{dr^2}{g_{\rm eff}^2r^2}.
\end{eqnarray}
The boundary CFT metric thus has a redshift factor
\begin{equation}
\label{redshift}
\lambda=\fft{g_{\rm eff}}g,
\end{equation}
which may be removed by rescaling asymptotic time
\begin{equation}
t \rightarrow t' = \lambda t \, .
\end{equation}
Thus, in the CFT, all thermodynamic quantities in this section ought to be
rescaled via
\begin{equation}
\{E, T, \Phi, \Omega\} \rightarrow \frac{1}{\lambda} \left\{E, T, \Phi, \Omega \right\} .
\end{equation}
We will only perform the scaling explicitly for the energy, since it is
the quantity which plays a key role in the discussion of the mass to
charge relation.

\section{The weak gravity conjecture and $M/Q$}
\label{MassToChargeSection}

It is not surprising that the relation between the mass $m$ and the charge
$q$ of extremal black hole solutions is modified in the presence of curvature
corrections. In light of the weak gravity conjecture, which emerged from the
ideas explored in \cite{Vafa:2005ui} and later refined in
\cite{ArkaniHamed:2006dz}, it is interesting to examine the precise
dependence of the mass on the $R$-charge for the solutions we have
constructed above.

One of the key points emphasized in \cite{Vafa:2005ui} is the fact that string
theory, or any theory of quantum gravity, puts constraints on low energy
physics, so that not every (consistent) effective field theory can in fact
be UV completed.  Thus, the \emph{landscape} of ``good'' theories -- those
which are compatible with quantum gravity -- is much smaller than the actual
\emph{swampland} of all effective field theories which do not have
a UV completion.
Building on the simple observation that ``gravity is the weakest force,''
the authors of \cite{ArkaniHamed:2006dz} conjectured that there should always
be elementary objects whose mass to charge ratio is smaller than the
corresponding one for macroscopic extremal black holes. The presence of such
objects would then provide a decay channel for extremal black holes,
alleviating the problem of remnants.
Thus, according to the weak gravity conjecture, the mass/charge relation $m=q$
for extremal black holes cannot be exact, but must instead receive corrections
as the charge $q$ decreases. Furthermore, the deviation from the extremal limit
is expected to become more pronounced as the charge becomes smaller.

An analysis of higher derivative corrections to the mass/charge ratio of
four-dimensional, \emph{asymptotically flat} black holes was performed in
\cite{Kats:2006xp}. In the examples where the sign of the correction to
$m/q$ could be verified from UV physics, it was found to be negative, in agreement with
the claims of \cite{ArkaniHamed:2006dz}. Similar results appeared more
recently \cite{Giveon:2009da} in the context of $d$-dimensional black holes
with two electric charges, which are solutions corresponding to fundamental
strings with generic momentum and winding on an internal circle.
While the weak gravity conjecture was originally phrased in terms of four-dimensional,
asymptotically flat black holes, it is worth exploring its analog in the
context of extremal black holes in AdS. In particular, there have been
suggestions in the literature that the correction to $m/q$ might be
somehow tied to the correction to the shear viscosity to entropy density
ratio $\eta/s$ (as well as to the charge conductivity)
\cite{Kats:2007mq,Myers:2009ij,Cremonini:2009sy}.
When discussing the effects of higher derivatives on various transport coefficients, the authors of \cite{Myers:2009ij} included an analysis of
$m/q$ for five-dimensional $R$-charged black holes, and their results were
in qualitative agreement with those of \cite{Kats:2006xp}.

Given our analysis in this paper, we may extend some of these studies to $R$-charged
solutions in $d$-dimensions.  As we will see, our results will be similar to
those already found in \cite{Kats:2006xp} and \cite{Myers:2009ij}.
Moreover, we emphasize that in five dimensions the deviation from the
linear extremal mass-charge relation predicted by the weak gravity conjecture
seems to be intimately tied to the corrections observed in some of the
hydrodynamic calculations in AdS$_5$/CFT$_4$.
Such a connection could be a consequence of gravity
constraining the set of allowed dual CFTs.

In Section \ref{ThermoSection} we extracted the energy of the corrected
$R$-charge solutions from the boundary stress tensor.  In this case, the
mass to charge ratio is given simply by
\begin{equation}
\frac{m}{q} = \frac{1}{\lambda}\frac{E}{\mathcal{Q}},
\end{equation}
where the energy $E$ is given in (\ref{eq:E}), but must be rescaled by the
redshift factor $\lambda$ introduced in (\ref{redshift}) to ensure proper
boundary asymptotics.  Recall that the normalized charge $\mathcal Q$ is
given by $\mathcal{Q} = (\omega_{d-2}/2\kappa_d^2) Q$.  Since we are
interested in $m/q$ for extremal black holes, we make use of the extremality
condition $T=0$ as well as the horizon condition $f(r_+)=0$.

Although we ultimately want to consider black holes in AdS, we start by
setting $g=0$ and $k=1$ in order to examine $m/q$ for asymptotically flat
solutions with a spherical horizon, as was done in \cite{Kats:2006xp}.
We find
\begin{eqnarray}
\frac{m}{q}&=&
\left(\frac{m}{q}\right)_0\bigg(1 - \frac{\ala}{r_+^2}\frac{(d-3)^2(d-4)^2}{2(d-2)(3d-7)} -\frac{\alb}{r_+^2}\frac{(d-3)^2(2d^2-11d+16)}{2(d-2)(3d-7)}  \nonumber \\
&& \kern6em -\frac{\alc}{r_+^2}\frac{(d-3)(2d^3-16d^2+45d-44)}{(d-2)(3d-7)}  \bigg) \, ,
\end{eqnarray}
where
\begin{equation}
\left(\frac{m}{q}\right)_0 = \sqrt{\frac{2(d-2)}{d-3}}
\end{equation}
is the uncorrected mass to charge ratio. Note first of all that, independent
of the number of dimensions, the correction is always negative whenever
the $\alpha_i$'s are positive.
Furthermore, as one can easily check by trading
$r_+$ dependence for $Q$ dependence, the $1/r_+^2$ factor in front of the
higher derivative corrections implies that the deviation from the linear
relation $m\sim q$ is enhanced as the charge decreases. This was precisely one
of the predictions of the weak gravity conjecture, and was also observed in
\cite{Kats:2006xp}. Of course to say anything more about the precise
form of the correction, one needs to determine the couplings.

The expressions corresponding to spherical horizon black holes in AdS are
significantly more complicated.  Here we quote the result in $d=5$, and
relegate the $d=4$ and $d=6$ cases to the appendix, since they are
qualitatively the same:
\begin{eqnarray}
\label{eq:mqads5}
\left(\frac{m}{q}\right)_{d=5}&=&
\left(\frac{m}{q}\right)_{0,d=5}\bigg(1 - \ala
\frac{(816\beta^3+1024\beta^2 +300\beta+1)}{6r_+^2(1+2\beta)(2+3\beta)}
\nonumber \\
&& - \alb \frac{(336\beta^3+392\beta^2 +132\beta+11)}{6r_+^2(1+2\beta)
(2+3\beta)} -\alc \frac{(564\beta^3+586\beta^2 +216\beta+31)}{6r_+^2(1+2\beta)
(2+3\beta)} \bigg), \qquad
\end{eqnarray}
where $\beta = g^2r_+^2$ and
\begin{equation}
\left(\frac{m}{q}\right)_{0,d=5} = \frac{\sqrt{3}(2+3\beta)}{2\sqrt{1+2\beta}}.
\end{equation}
As in the asymptotically flat case, the corrections are sensitive to the sign
of the couplings, and will necessarily push the solution below the extremal
limit when all the $\alpha_i$ are positive.  Of course, if some of the
couplings are negative the various terms can conspire to yield a positive
correction to the mass to charge ratio.   However, if the weak gravity
conjecture holds, we would expect that, in an  effective theory that is
consistent with gravity in the UV, the couplings would be  constrained in
such a way as to lower $m/q$.   Again, this underlines the importance of
obtaining the higher derivative couplings from UV physics.

In the asymptotically Minkowski case we observed that $m/q$ became smaller
as the charge decreased, since the overall $1/r_+^2$ factor decreases
monotonically as $r_+$ increases.  Here the AdS black hole situation is similar
only as long as $r_+$ does not become too large.  When $r_+ \sim 1/g$, the
coefficient of the $\alpha_3$ term reaches a minimum and starts growing as
$r_+$ increases. This effect was already noticed in \cite{Myers:2009ij} and
is intrinsic to the AdS geometry -- it reflects the fact that the size of the
black hole is becoming of the same order as the $AdS$ radius.

One of the results of the investigations of the hydrodynamic regime of
four-dimensional SCFTs has been the universality \cite{Buchel:2003tz}
of the shear viscosity to
entropy ratio, $\eta/s = 1/4\pi$ in the leading supergravity approximation.
Studies of $R^4$ corrections \cite{Buchel:2004di,Buchel:2008sh,Myers:2008yi}
increased the ratio, and seemed to favor
the existence of a new bound in nature, $\eta/s \geq 1/4\pi$, the celebrated
KSS bound. However, with the inclusion of curvature-squared corrections the
bound has been shown to be violated by $1/N$ effects on the CFT side
\cite{Kats:2007mq,Buchel:2008vz,Myers:2009ij,Cremonini:2009sy}.
The size of the violation is related to the two central charges $a$, $c$ of
the dual four-dimensional CFT. Holographic Weyl anomaly matching demonstrates
that the coefficient of the $R^2$ terms in the action is proportional to
$(c-a)/c$, and it is precisely the quantity $c-a$ which controls the strength
of the correction to $\eta/s$, with $c-a>0$ necessarily giving violation of
the bound.

Until recently, all the available CFT examples with a known gravity dual
corresponded to $c-a>0$, so that violating the $\eta/s$ bound seemed to be
the rule rather than the exception. However, a large class of four
dimensional $\mathcal N =2$ CFTs was constructed recently in
\cite{Gaiotto:2009we}, and shown in \cite{Gaiotto:2009gz} to contain
examples with $c-a<0$ and a known dual gravitational description.
These are quiver gauge theories which can be viewed as arising from M5 branes
wrapping a Riemann surface. Furthermore, one can add non-compact branes
that intersect the surface at points (punctures on the Riemann surface).
In the large $N$ limit, these yield a large class of AdS$_5$ compactifications
of M-theory with four-dimensional $\mathcal N =2$ supersymmetry, some of which
correspond to $c-a<0$.

In light of these constructions, the requirement of $c-a>0$ which seemingly
arises from the weak gravity conjecture is rather puzzling.  Ideally, we
may have expected the gravity duals to restrict the set of allowed CFTs,
effectively placing the ones with $c-a<0$ into the swampland.  However, such
a statement would have to be reconciled with the results of
\cite{Gaiotto:2009gz}, which found no such sign restrictions.  
Still, it is remarkable that the issue of the sign of $c-a$
arises not only in the computation of transport coefficients, but also in the
context of the weak gravity conjecture.  We illustrate this connection with
a simple example.

To make contact with the AdS/CFT work on transport coefficients, we take
$d=5$ and consider the Weyl-tensor-squared corrected action
\beq
\label{eq:Csq}
S_{\rm bulk}=-\fft1{2\kappa_5^2}\int_{\mathcal M}
d^{5}x\sqrt{-g}\, \Biggl[R-\frac{1}{4}F^2+12 g^2 +
\alpha\left(\frac{1}{6}R^2-\frac{4}{3} R_{\mu\nu}R^{\mu\nu}+ R_{\mu\nu\rho\sigma}R^{\mu\nu\rho\sigma}\right)\Biggr].
\eq
This choice is motivated by the general form of the supersymmetric higher
derivative action that was used in \cite{Cremonini:2009sy} to obtain the
corrections to $\eta/s$ in $\mathcal N=1$ SCFT.  In fact, $\eta/s$ for
(\ref{eq:Csq}) can be read off from \cite{Cremonini:2009sy},
and takes the form
\beq
\frac{\eta}{s}=\frac{1}{4\pi}[1-4\alpha (2-q)],
\eq
where $0\leq q \leq 2$, and $q$ is the $R$-charge in the notation of
\cite{Cremonini:2009sy}.
The main feature to point out is that, since $(2-q)$ is non-negative,
the condition $\alpha>0$ (or alternatively $c-a>0$) always leads to violation of the $\eta/s$ bound, and also guarantees that the entropy increases.
But $\alpha>0$ is also the requirement needed to satisfy the weak gravity conjecture.
In fact, for the Weyl squared correction in $d=5$, our result for $m/q$ reduces to:
\beq
\left(\frac{m}{q}\right)_{d=5} = \left(\frac{m}{q}\right)_0 \biggl(1-\alpha
\frac{168\beta^3+156\beta^2 +60\beta+11}{4 \, r_+^2 \, (1+2\beta)(2+3\beta)} \biggr)\, .
\eq
While here we have focused on five dimensions, these features are generic in other dimensions as well (as can be inferred from the $m/q$ expressions in the
appendix).

For a less trivial example in $d=5$ we can look at the most general
four-derivative action describing $R$-charged solutions, which has
been studied in \cite{Myers:2009ij,Cremonini:2009sy}, and can be
reduced -- via appropriate field redefinitions -- to the simple form:
\beq
e^{-1}\delta \mathcal{L} = c_1 R_{\mu\nu\rho\sigma}R^{\mu\nu\rho\sigma}+
c_2 R_{\mu\nu\rho\sigma}F^{\mu\nu}F^{\rho\sigma} + c_3 (F^2)^2 +c_4 F^4
+ c_5 \epsilon^{\mu\nu\rho\sigma\lambda} A_\mu R_{\nu\rho\alpha\beta}R_{\sigma\lambda}^{\;\;\;\;\alpha\beta} \, .
\eq
The effect of such terms on the shear viscosity to entropy density ratio can
be read off from \cite{Myers:2009ij,Cremonini:2009sy}, and for the special
case where the terms are constrained by supersymmetry (so that all the
$c_i$'s are related to each other), one finds:
\beq
\frac{\eta}{s}=\frac{1}{4\pi} \Bigl[ 1-c_1 \,g(Q)\Bigr] \, ,
\eq
where $g(Q)$ is a non-negative function of $R$-charge. The mass to charge
relation for this case has been worked out in \cite{Myers:2009ij} and
exhibits the same behavior we found in the simpler Weyl-tensor-squared case:
\beq
\left(\frac{m}{q}\right)_{d=5} = \left(\frac{m}{q}\right)_0 \biggl(1-c_1 \,f(r_+) \biggr)\, ,
\eq
where again $f(r_+)$ is always positive. While the precise form of the
corrections to $m/q$ and $\eta/s$ is different, the behavior required by
the weak gravity conjecture (in this case $c_1>0$) is again correlated
with the violation of the viscosity to entropy bound.

The correlation between the behavior of $\eta/s$ and the corrections to
$m/q$ is intriguing. It hints, at least in the five-dimensional context, at
a close connection between the sign of $c-a$ and possible fundamental
constraints on the gravitational side of the duality. However, in this case
one would need to understand the role played by the strongly coupled theories
investigated in \cite{Gaiotto:2009gz}, which allow for negative $c-a$.
We should also point out that studies of causality in the CFT
\cite{Brigante:2008gz,Buchel:2009tt} as well as the requirement of positive
energy measurements in collider experiments \cite{Hofman:2008ar,Hofman:2009ug}
(also note the work of \cite{Shapere:2008zf}) have resulted in bounds on the
central charges $a$ and $c$, but so far have not lead to any restrictions on
the actual sign of $c-a$. Nevertheless, theories with $c-a<0$ would naively
seem to be in conflict with the weak gravity conjecture, and thus may be
expected to possess unusual features. We note that these ideas have already
been explored in several contexts. For example, \cite{Adams:2006sv} have
identified consistency conditions for effective field theories with
a UV completion, based on the idea that the signs of certain higher
dimensional operators must be strictly positive. Such arguments, however, still
need to be fully extended to generic gravitational settings.

Having a geometrical understanding of the origin of the higher derivative
couplings -- and of their sign in particular -- would also be valuable.
For example, for the case of ungauged $\mathcal{N}=2$, $d=5$ supergravity
(obtained by reducing $d=11$ supergravity on a $CY_3$), the coupling
of the $R_{\mu\nu\rho\sigma}R^{\mu\nu\rho\sigma}$ term can be shown to be
related to the second Chern class of the $CY_3$, which is known to be positive.
For the case of $\mathcal{N}=2$, $d=5$ gauged supergravity (which is needed to
discuss black holes in AdS), the compactification manifold would be a
five-dimensional Sasaki-Einstein manifold, and the geometric origin of the
higher derivative couplings is less clear. While there is work
\cite{Gabella:2009ni,Gabella:2009wu} relating geometric data of generic
supersymmetric AdS$_5$ solutions of type IIB supergravity to the central
charges $a$, $c$ of the dual CFTs, so far this applies only to the leading
supergravity approximation, where $a=c=\mathcal{O}(N^2)$. Thus, it would be
valuable to generalize these constructions to accommodate finite $N$
corrections to the central charges. Whether through geometric data, or through
consistency arguments on the field theory side, a better understanding of the
signs of the higher derivative gravitational couplings is needed. This is
especially relevant if we want to achieve a deeper insight into the weak
gravity conjecture, and how it is tied to seemingly unrelated quantities such
as hydrodynamic transport coefficients.

\acknowledgments

S.C. would like to thank the Michigan Center for Theoretical Physics
for their hospitality during the final stages of this project.
This work was supported in part by the US Department of Energy under grant
DE-FG02-95ER40899. The work of S.C. has been supported by the
Cambridge-Mitchell Collaboration in Theoretical Cosmology, and the Mitchell
Family Foundation.

\appendix
\section{The mass to charge ratio in AdS}

For the case of asymptotically AdS solutions with a flat boundary,
i.e. $k=0$, $g\neq0$, we find that the
mass to charge ratio is:
\begin{eqnarray}
\frac{m}{q}&& = \left(\frac{m}{q}\right)_0\bigg(1 - \tilde{\alpha}_1 g^2\frac{(d-1)(7d^3-27d^2+8d+32)}{2(d-2)(3d-7)}  \nonumber \\
&& \kern8em - \tilde{\alpha}_2 g^2\frac{(d-1)(2d^3-3d^2-19d+32)}{2(d-2)(3d-7)} - \alc g^2\frac{(d-3)(d-4)}{2}  \bigg) \nonumber \\
&&= \left(\frac{m}{q}\right)_0\bigg(1 - \ala g^2\frac{(d-1)(7d^3-27d^2+8d+32)}{2(d-2)(3d-7)} \nonumber \\
&& \kern8em - \alb g^2\frac{(d-1)(2d^3-3d^2-19d+32)}{2(d-2)(3d-7)} \nonumber \\
&& \kern8em -\alc g^2\frac{(2d^4-10d^3+21d^2-37d+36)}{(d-2)(3d-7)} \bigg)\,,
\end{eqnarray}
where
\begin{equation}
\left(\frac{m}{q}\right)_0 = gr_+\sqrt{\frac{2(d-2)^3}{(d-1)(d-3)^2}}\, .
\end{equation}
We note that if the redshift factor $\lambda$ had not been taken into account,
the correction to the mass/charge ratio for the $k=0$ Gauss-Bonnet term ($\tilde{\alpha}_1=0, \tilde{\alpha}_2=0$)
would have vanished. It is precisely
the addition of the redshift factor which is responsible for generating the correction.

For the case of asymptotically AdS solutions with a spherical horizon, i.e. $k=1$, $g \neq 0$, the expressions are rather more complicated.  For
$d=4$, we find
\begin{eqnarray}
\left(\frac{m}{q}\right)_{d=4}&& = \left(\frac{m}{q}\right)_0\bigg(1 - 12\tilde{\alpha}_1 g^2  - \tilde{\alpha}_2 \frac{(54\beta^2 +21\beta+1)}{5r_+^2(1+2\beta)} \bigg) \nonumber \\
&&=  \left(\frac{m}{q}\right)_0\bigg(1 - 12 \ala g^2 - \alb \frac{(54\beta^2 +21\beta+1)}{5r_+^2(1+2\beta)}  -4 \alc \frac{(24\beta^2 +6\beta+1)}{5r_+^2(1+2\beta)} \bigg)\, ,  
\end{eqnarray}
where $\beta = g^2r_+^2$, and
\begin{equation}
\left(\frac{m}{q}\right)_{0,d=4} = \frac{2(1+2\beta)}{\sqrt{1+3\beta}}\, .
\end{equation}

For $d=5$, we have
\begin{eqnarray}
\left(\frac{m}{q}\right)_{d=5} =&& \left(\frac{m}{q}\right)_0\bigg(1 - \tilde{\alpha}_1 \frac{(816\beta^3+1024\beta^2 +300\beta+1)}{6r_+^2(1+2\beta)(2+3\beta)}  \nonumber \\
&& - \tilde{\alpha}_2 \frac{(336\beta^3+392\beta^2 +132\beta+11)}{6r_+^2(1+2\beta)(2+3\beta)} - \alc \frac{(3\beta^2 + 2 \beta-2)}{r_+^2(2+3\beta)} \bigg) \nonumber \\
=&&  \left(\frac{m}{q}\right)_0\bigg(1 - \ala \frac{(816\beta^3+1024\beta^2 +300\beta+1)}{6r_+^2(1+2\beta)(2+3\beta)} \nonumber \\
&& - \alb \frac{(336\beta^3+392\beta^2 +132\beta+11)}{6r_+^2(1+2\beta)(2+3\beta)} -\alc \frac{(564\beta^3+586\beta^2 +216\beta+31)}{6r_+^2(1+2\beta)(2+3\beta)} \bigg), \qquad
\end{eqnarray}
where
\begin{equation}
\left(\frac{m}{q}\right)_{0,d=5} = \frac{\sqrt{3}(2+3\beta)}{2\sqrt{1+2\beta}}.
\end{equation}
This result corresponds to (\ref{eq:mqads5}) given in
Section~\ref{MassToChargeSection}.

Similarly, for $d=6$:
\begin{eqnarray}
\left(\frac{m}{q}\right)_{d=6} =&& \left(\frac{m}{q}\right)_0\bigg(1 -1 \tilde{\alpha}_1 \frac{(15500\beta^3+23445\beta^2+8325\beta+81)}{22r_+^2(3+4\beta)(3+5\beta)} \nonumber \\
&& - \tilde{\alpha}_2 \frac{(275\beta^2 + 195\beta+27)}{4r_+^2(3+5\beta)} - 3 \alc \frac{(20\beta^3+27\beta^2 -7\beta-9)}{r_+^2(3+4\beta)(3+5\beta)} \bigg) \nonumber \\
= && \left(\frac{m}{q}\right)_0\bigg(1 - \alpha_1 \frac{(15500\beta^3+23445\beta^2+8325\beta+81)}{22r_+^2(3+4\beta)(3+5\beta)}  \nonumber \\
&& - \alpha_2 \frac{(275\beta^2 + 195\beta+27)}{4r_+^2(3+5\beta)}  - 3 \alc \frac{(3340\beta^3+4549\beta^2 + 2153\beta+369)}{22r_+^2(3+4\beta)(3+5\beta)} \bigg) \, , 
\end{eqnarray}
where
\begin{equation}
\left(\frac{m}{q}\right)_{0,d=6} = \frac{2\sqrt{2}(3+4\beta)}{3\sqrt{3+5\beta}}\, .
\end{equation}
A general $d$-dimensional expression may be obtained in principle, although
it is not expected to be particularly illuminating.


\end{document}